\documentclass[doublecol]{epl2}

\title{Supersymmetrical Separation of Variables for Scarf II Model: Partial Solvability}

\author{M.V. Iof\/fe\inst{1}\footnote{E-mail: m.ioffe@pobox.spbu.ru}   \and E.V. Krupitskaya\inst{1}\footnote{E-mail: e.v.krup@yandex.ru}
\and D.N. Nishnianidze\inst{1,2}\footnote{E-mail: cutaisi@yahoo.com}
}

\institute{
  \inst{1} Saint-Petersburg State University,
198504 St.-Petersburg, Russia\\
  \inst{2} Akaki Tsereteli State University, 4600 Kutaisi, Republic of Georgia
}

\abstract{
Recently, a new quantum model - two-dimensional generalization of the Scarf II -
was completely solved analytically by SUSY method for the integer values of parameter. Now, the same integrable model, but with arbitrary values
of parameter, will be studied by means of supersymmetrical intertwining relations. The Hamiltonian does not allow the conventional separation
of variables, but the supercharge operator does allow, leading to the partial solvability of the model. This approach, which can be called as the first variant of SUSY-separation, together with shape invariance of the model, provides analytical calculation of the part of spectrum and corresponding wave functions (quasi-exact-solvability). The model is shown to obey two different variants of shape invariance which can be combined effectively in construction of energy levels and wave functions.
}

\begin{document}

\maketitle

\newcommand{\be}{\begin{equation}}
\newcommand{\ee}{\end{equation}}
\newcommand{\ba}{\begin{eqnarray}}
\newcommand{\ea}{\end{eqnarray}}
\renewcommand\thefootnote{\alph{footnote}}

\section{Introduction}

Supersymmetrical Quantum Mechanics gave a new impetus to development of new analytical methods in study of different quantum models \cite{witten} -  \cite{fernandez-review}. The main ingredients and results of SUSY approach in Quantum Mechanics are isospectral (or, almost isospectral) quantum systems and supersymmetrical intertwining relations \cite{witten}, shape invariance \cite{gendenstein} - \cite{new}, higher order supercharges \cite{ais}, supersymmetry of multidimensional \cite{abi} - \cite{david2}, multiparticle \cite{ghosh} and matrix \cite{pauli} quantum systems.

One of the directions of research within the framework of this approach is investigation of two-dimensional quantum models which are not amenable to standard \cite{miller} separation of variables. Recently, two different variants of supersymmetrical (SUSY) separation of variables were proposed \cite{new}, \cite{ioffe1} - \cite{physrev} for analysis of such two-dimensional quantum systems. Both procedures are based on supersymmetrical intertwining relations \cite{witten}, \cite{abi} with second order supercharges \cite{david} - \cite{david2}, \cite{ioffe1} and shape invariance \cite{gendenstein}, \cite{krive}, \cite{new}, \cite{ioffe1}, \cite{mallow}, \cite{shape2011} property. The first variant is applicable if the intertwining operator (supercharge) allows the standard separation of variables as it happens for supercharges with Lorentz metric. It was used in practice for two-dimensional generalizations of Morse \cite{new}, \cite{ioffe1}, P\"oschl-Teller \cite{IV}, and periodic Lame \cite{periodic} models. The second variant works if one of the intertwined Hamiltonians allows the standard separation due to specific choice of parameters, but its superpartner still does not. Such situation gives a chance to solve the problem completely - to find
energies and wave functions of all bound states. This procedure was applied recently to Morse \cite{physrev}, P\"oschl-Teller \cite{INV} models, and quite recently \cite{IKN}, to the new model - two-dimensional generalization of Scarf II model. In the latter case, one of intertwined Hamiltonians is separable if one of parameters takes only negative integer values $a=-k.$

In the present paper, the same two-dimensional Scarf II model but with arbitrary values of parameter $a$ will be studied by means of the first procedure of SUSY separation of variables. This method provides a part of energy spectrum and corresponding wave functions in analytical form. Thus, the model belongs to the class of partially (i.e. quasi-exactly-solvable) models - the intermediate class between completely (exactly solvable) and analytically unsolvable models. This class was considered starting from 80-ties \cite{razavy} - \cite{kamran}. In particular, the elegant algebraic method of construction of one-dimensional quasi-exactly-solvable (and sometimes, of exactly solvable) quantum models was elaborated in \cite{ushveridze} - \cite{shifman2}. This approach is applicable to two-dimensional problems as well, but only in curved spaces with nontrivial metrics \cite{shifman1}. The approach of the present paper allows to study the different class of quantum models which do not allow standard separation of variables. It is necessary to notice that by construction these models are integrable with the symmetry operators of fourth order in momenta.
In Section 2, the zero modes of supercharge will be found, and their linear combinations, which are the eigenfunctions of Hamiltonian, will be built. In Section 3, the shape invariance of the model will be used to enlarge the variety of known wave functions. It will be shown that the second shape invariance exists for this model leading to additional relations between different wave functions.

\section{Wave functions in subspace of zero modes of $Q^+$}
\subsection{Formulation of the model}

The main tools of the supersymmetrical approach in two-dimensional Quantum Mechanics are the supersymmetrical intertwining relations:
\be
H^{(1)}Q^{+}=Q^{+}H^{(2)}; \qquad Q^{-}H^{(1)}=H^{(2)}Q^{-}, \label{int}
\ee
for two partner two-dimensional Hamiltonians of Schr\"odinger form
\ba
& &H^{(i)}=-\Delta^{(2)}+V^{(i)}(\vec x);\,\, i=1,2;\quad \vec x = (x_1, x_2);\label{H}\\
& &\quad \Delta^{(2)}\equiv \partial_1^2+\partial_2^2;\quad \partial_i\equiv \frac{\partial}{\partial x_i} \nonumber
\ea
with mutually conjugate supercharges $Q^{\pm}$ of second order in derivatives. First of all, one has to find solutions of Eq.(\ref{int}), i.e. to find such potentials $V^{(1, 2)}(\vec x)$ and such coefficient functions of second order supercharges $Q^{\pm}$ that (\ref{int}) are fulfilled.
In terms of these unknown functions, we deal with a complicate system of nonlinear differential equations of second order. Due to suitable choice of ansatzes,   a list of particular solutions of (\ref{int}) was found \cite{david}, \cite{ioffe1}. A part of these solutions was shown to allow the analytical construction of spectra and wave functions: depending on chosen values of parameters, the partial \cite{new}, \cite{IV}, \cite{periodic}, \cite{ioffe1} and/or complete \cite{physrev},  \cite{INV} solutions of corresponding models were obtained.

Recently, new results \cite{mallow} obtained in one-dimensional shape invariance allowed to find also new solutions \cite{shape2011} of (\ref{int}) in two-dimensional framework. In particular, the two-dimensional generalization of Scarf II model was among these new solutions, and just this system will be considered below. The potentials\footnote{Slightly different notations for coupling constants are chosen here as compared with \cite{IKN}}:
\ba
&V^{(1),(2)}=-2\lambda^{2}a(a\mp1)(\frac{1}{\cosh^{2}(\lambda x_{+})}-\frac{1}{\sinh^{2}(\lambda x_{-})})+
\nonumber\\
&+\frac{2B(A+\lambda)\sinh(2\lambda x_{1})+(B^2-A^2-2A\lambda)}{\cosh^{2}(2\lambda x_{1})}+\nonumber\\
&+\frac{2B(A+\lambda)\sinh(2\lambda x_{2})+(B^2-A^2-2A\lambda)}{\cosh^{2}(2\lambda x_{2})}, \label{V}
\ea
and the second order supercharges:
\ba
&Q^{+}=(Q^{-})^{\dag}=4\partial_{+}\partial_{-}+4a\lambda\tanh (\lambda x_{+})\partial_{-}+\nonumber\\
&+4a\lambda\coth (\lambda x_{-})\partial_{+}+4a^{2}\lambda^2\tanh (\lambda x_{+})\coth (\lambda x_{-})-\nonumber\\
&-\frac{2B(A+\lambda)\sinh( 2\lambda x_{1})+(B^2-A^2-2A\lambda)}{\cosh^{2}( 2\lambda x_{1})}+\nonumber\\
&+\frac{2B(A+\lambda)\sinh( 2\lambda x_{2})+(B^2-A^2-2A\lambda)}{\cosh^{2}(2\lambda x_{2})}, \label{Q}
\ea
solve the intertwining relations (\ref{int})($x_{\pm}\equiv x_{1} \pm x_{2}, \partial_{\pm}=\partial / \partial x_{\pm}; \,\, \lambda , a$ are real parameters, and $A,\,B >0).$ As is typical to the approach, both potentials (\ref{V}) are not amenable to standard separation of variables, but they correspond to integrable Hamiltonians (\ref{H}) with symmetry operators of fourth order in derivatives:
$R^{(1)}=Q^+Q^-, \,\, R^{(2)}=Q^-Q^+.$

Quite recently, it was proven that this model is completely (exactly) solvable for the values of parameter $a=-k.$
The whole spectrum of bound states and wave functions were found analytically by means of intertwining relations (\ref{int}) and shape invariance of (\ref{V}) under the change $a \to a-1$ (see Section 3 below).

\subsection{Zero modes of supercharge $Q^+$}

The same model will be studied below for arbitrary values of $a$ in the framework of SUSY separation of variables of the first kind (see \cite{new}, \cite{IV}, \cite{periodic}, \cite{ioffe1}). The basic idea is that the supercharge $Q^{+}$ for Lorentz (hyperbolic) form of metric, like in (\ref{Q}), is amenable to conventional separation of variables \cite{new}, \cite{ioffe1}. Indeed, after a suitable similarity transformation:
\ba
&Q^+=\exp{(\chi )} q^+ \exp{(-\chi )};\nonumber\\
&\chi(\vec x)=-a\ln|\cosh(\lambda x_+)\sinh(\lambda x_-)|,\nonumber
\ea
we obtain the operator with separated variables:
\ba
&q^+= \partial_1^2-\partial_2^2 - f(x_1) + f(x_2); \nonumber\\
&f(x)= \frac{2B(A+\lambda)\sinh(2\lambda x)+(B^2-A^2-2A\lambda)}{\cosh^{2}(2\lambda x)}\label{f}
\ea
Therefore, depending on explicit form of $f(x),$ we get a chance to find analytically the zero modes $\Omega_n(\vec x)$ of supercharge $Q^+:$
\ba
&\Omega_n(\vec x)=
\exp{(\chi(\vec x))}\omega_n(\vec x)=\label{zero}\\
&=|\cosh(\lambda x_+)\sinh(\lambda x_-)|^{-a}\omega_n(\vec x);\, q^+\omega_n(\vec x)=0.\nonumber
\ea
The zero modes $\omega_n(\vec x)=\eta_n(x_1)\eta_n(x_2)$ of operator $q^+$ are expressed in terms of solutions of one-dimensional Schr\"odinger equations:
\be
\biggl(-\partial_1^2 + f(x_1)\biggr)\eta_n(x_1) = \epsilon_n\eta_n(x_1). \label{eta}
\ee
Fortunately, for the model under consideration, function $f(x)$ is such that this one-dimensional model belongs to the class of exactly solvable models \cite{infeld}, \cite{dabrowska}.  It is known as Scarf II (hyperbolic Scarf) systems \cite{scarf}, and its spectrum and wave functions were built analytically in terms of Jacobi polynomials \cite{jacobi}:
\ba
&\eta_{n}(x)=i^{n}(\cosh(2\lambda x))^{-A/2\lambda}\cdot\label{eta-epsilon}\\
&\cdot\exp[-(B/2\lambda)\arctan\sinh(2\lambda x)]P_{n}^{(\gamma, \beta)}(i\sinh(2\lambda x));\nonumber\\
&\gamma \equiv  -(A/2\lambda + iB/2\lambda + 1/2);\, \beta \equiv \gamma^{\star};\,\epsilon_{n}=-(A - 2n\lambda )^{2}. \nonumber
\ea
Therefore, all normalizable
zero modes $\Omega_n(\vec x)$ (and their components $\omega_n$) are known and can be used to find the wave functions of the Hamiltonian $H^{(2)}.$

It follows \cite{new}, \cite{ioffe1} directly from the intertwining relations (\ref{int}) that the variety of zero modes is closed under the action of operator $H^{(2)}.$ In other words, the action of $H^{(2)}$ onto $\Omega_n$ gives the linear combination of $\Omega'$s:
\be
H^{(2)}\Omega_n(\vec x)=\sum_{k=0}^{N}c_{nk}\Omega_k(\vec x),
\label{C}
\ee
where coefficients $c_{nk}$ form the $(N+1)\times (N+1)$ constant matrix $\widehat C.$ Diagonalization of this matrix, if possible, will provide both wave functions and energy eigenvalues of $H^{(2)},$ though not all, in general.

\subsection{Calculation of matrix $\widehat C$}

From this point and below, we'll simplify our presentment by choosing the specific value $\lambda =1/2$ in all formulas above. It is more convenient to look for the explicit form of matrix $\widehat C$ in terms of $\omega_n,$ replacing $H^{(2)}$ by its similarity transform:
\ba
&h^{(2)}\equiv \exp{(-\chi )}H^{(2)}\exp{(+\chi )}=\label{h}\\
&=-(\partial_1^2+\partial_2^2)+f(x_1)+f(x_2)+2a\tanh\frac{x_+}{2}\, \partial_++\nonumber\\
&+2a\coth\frac{x_-}{2}\, \partial_--a^2 ; \nonumber\\
&h^{(2)}\omega_n(\vec x)=\sum_{k=0}^{N}c_{nk}\omega_k(\vec x).\nonumber
\ea
Since the functions $\omega_n$ are factorized onto multipliers $\eta_n(x_1),\,\eta_n(x_2)$ (see (\ref{eta})), the action of $h^{(2)}$ can be simplified:
\ba
&h^{(2)}\omega_{n}(\vec x)=\biggl[2\epsilon_{n}-a^2+\label{homega}\\
&+\frac{2a}{\sinh x_{1}-\sinh x_{2}}\biggl(\cosh x_{1}\, \partial_{1}-\cosh x_{2} \,\partial_{2}\biggr)\biggr]\omega_{n}.
\nonumber
\ea
Introducing new variables $z_{1}=i \sinh x_{1};\quad z_{2}=i \sinh x_{2},$
we consider separately a part of (\ref{homega}):
\ba
&\biggl(\cosh x_{1}\, \partial_{1}-\cosh x_{2}\, \partial_{2}\biggr)\omega_{n}=
i(-1)^n\omega_0\cdot\label{omega-pi}\\
&\cdot\biggl[\biggl(z_{1} -z_{2} \biggr) A\, P_{n}^{(\gamma,\beta)}(z_{1}) P_{n}^{(\gamma,\beta)}(z_{2})
+ \Pi(z_{1}, z_{2})\biggr]. \nonumber
\ea
The function $\Pi$ appeared in (\ref{omega-pi}) can be written as:
\ba
&\Pi(z_1, z_2) \equiv (1-z_1^2)\partial_{z_1} P_{n}^{(\gamma,\beta)}(z_1) P_{n}^{(\gamma,\beta)}(z_2)-\nonumber\\
&-(1-z_2^2)P_{n}^{(\gamma,\beta)}(z_1) \partial_{z_2} P_{n}^{(\gamma,\beta)}(z_2), \nonumber
\ea
and by means of relations 22.8.1 of \cite{abramovich} between Jacobi polynomials $P_n(z_i)$ and their derivatives
over argument, it can be transformed as follows:
\ba
&\Pi(z_1,z_2)=-n(z_1-z_2)P_n^{(\gamma,\beta)}(z_1)P_n^{(\gamma,\beta)}(z_2)+\nonumber\\
&+\frac{2(n+\gamma)(n+\beta)}{2n+\gamma+\beta}\biggl(P_{n-1}^{(\gamma,\beta)}(z_1)P_n^{(\gamma,\beta)}(z_2)-\nonumber\\
&-P_n^{(\gamma,\beta)}(z_1)P_{n-1}^{(\gamma,\beta)}(z_2)\biggr).\label{2}
\ea
The expression in brackets is transformed by Christoffel-Darboux formula 22.12.1 \cite{abramovich}:
\ba
&P_{n-1}^{(\gamma,\beta)}(z_1)P_n^{(\gamma,\beta)}(z_2)-
P_n^{(\gamma,\beta)}(z_1)P_{n-1}^{(\gamma,\beta)}(z_2)=\label{3}\\
&=(z_2-z_1)\frac{k_nh_{n-1}}{k_{n-1}}\sum_{m=0}^{n-1}\frac{1}{h_m}
P_m^{(\gamma,\beta)}(z_1)P_m^{(\gamma,\beta)}(z_2),\nonumber
\ea
where the constants $h_n,\, k_n$ for Jacobi polynomials are defined in \cite{abramovich} and \cite{jacobi}:
\ba
&h_n=2^{\gamma +\beta +1}\frac{\Gamma (n+\gamma +1) \Gamma (n+\beta +1)}{(2n+\gamma +\beta +1)\Gamma (n+\gamma +\beta +1)};\nonumber\\
&k_n=2^{-n}\frac{(2n+\gamma +\beta)!}{n!(n+\gamma +\beta )!}. \label{hh}
\ea
Inserting (\ref{2}) and (\ref{3}) into (\ref{omega-pi}) we obtain:
\ba
&(\cosh x_1\,\partial_1-\cosh x_2\,\partial_2)\omega_n=\nonumber\\
&=i(-1)^n\omega_0\,(z_1-z_2)\biggl[ (A-n)
P_n^{(\gamma,\beta)}(z_1)P_n^{(\gamma,\beta)}(z_2)-\nonumber\\
&-\frac{2(n+\gamma)(n+\beta)}{2n+\gamma+\beta}\,\frac{k_nh_{n-1}}{k_{n-1}}\sum_{m=0}^{n-1}\frac{1}{h_m}
P_m^{(\gamma,\beta)}(z_1)P_m^{(\gamma,\beta)}(z_2)\biggr].\nonumber
\ea
Therefore, the relation (\ref{homega}) takes the form:
\ba
&h^{(2)}\omega_n=(2\epsilon_n-a^2)\omega_n-2a(-1)^n\omega_0\cdot\nonumber\\
&\cdot\biggl[ (A-n)P_n^{(\gamma,\beta)}(z_1)P_n^{(\gamma,\beta)}(z_2)-\nonumber\\
&-\frac{2(n+\gamma)(n+\beta)}{2n+\gamma+\beta}\,\frac{k_nh_{n-1}}{k_{n-1}}\sum_{m=0}^{n-1}\frac{1}{h_m}
P_m^{(\gamma,\beta)}(z_1)P_m^{(\gamma,\beta)}(z_2)\biggr].\nonumber
\ea
and taking into account, that $\omega_0 P_j^{(\gamma,\beta)}(z_1)P_j^{(\gamma,\beta)}(z_2)=(-1)^j\omega_j,$
it can be rewritten as:
\ba
&h^{(2)}\omega_n=\biggl(2\epsilon_n-a^2-2a(A-n)\biggr)\omega_n-\label{7}\\
&-\frac{4a(n+\gamma)(n+\beta)}{2n+\gamma+\beta}\,\frac{k_nh_{n-1}}{k_{n-1}}
\sum_{m=0}^{n-1}\frac{(-1)^{n+m}}{h_m}\omega_m.\nonumber
\ea
Thus, according to definition (\ref{C}) the matrix elements of $\widehat C$ are:
\ba
&c_{n,m}=0,\, for \, m>n;\,\,c_{n,n}=2\epsilon_n-a^2-2a(A-n);\nonumber\\
&c_{n,m}=-\frac{2ak_nh_{n-1}(-1)^m}{k_{n-1}h_m},\, for \, m<n;\,\, .\nonumber
\ea
Similarly to the models with Morse \cite{new}, \cite{ioffe1} and P\'oschl-Teller \cite{IV} potentials, matrix $\widehat C$ is triangular, and its diagonal elements give immediately the energy eigenvalues of $H^{(2)}:$
\be
E_n^{(2)}= 2\epsilon_n-a^2-2a(A-n)=-(A-n)^2-(A-n+a)^2.
\label{energies}
\ee

\subsection{Diagonalization of $\widehat C$}

In order to find the corresponding wave functions, one has to diagonalize the matrix $\widehat C,$ i.e. to find matrices $\widehat B$ and
$\widehat\Lambda=diag(\lambda_0, \lambda_1,...,\lambda_N),$ such that:
\be
\widehat B \widehat C = \widehat C \widehat\Lambda ;\quad\Leftrightarrow\quad \sum_{k=0}^{N}b_{ik}c_{kl}=\lambda_i b_{il}.
\label{diag}
\ee
The procedure of diagonalization is specific due to triagonality of $\widehat C,$ and the corresponding algorithm was proposed for the Morse model in \cite{new}. If this task will be solved, and matrix $\widehat B$ will be found, one can construct a variety of eigenfunctions of $H^{(2)}$ as linear combinations of zero modes $\Omega_n :$
\be
\Psi^{(2)}_{N-n}(\vec x) = \sum_{l=0}^{N} b_{nl}\Omega_l(\vec x).
\label{psiomega}
\ee

It is convenient to start an algorithm for solution of system of linear equations
(\ref{diag}) by solving the first line - with $i=0.$ Indeed, taking successively
$l=N,\, N-1, ..., l=0$ in (\ref{diag}), we obtain that $b_{0N}$ is arbitrary normalization
factor, and other $b_{0l}$ are found iteratively:
\ba
&b_{0(N-1)} = b_{0N}\frac{c_{N(N-1)}}{c_{NN}-c_{(N-1)(N-1)}};\nonumber\\
&b_{0(N-2)} = b_{0N}\biggl[\frac{c_{N(N-2)}}{c_{NN}-c_{(N-2)(N-2)}}+\nonumber\\
&+\frac{c_{N(N-1)}}{c_{NN}-c_{(N-1)(N-1)}}\cdot\frac{c_{(N-1)(N-2)}}{c_{NN}-c_{(N-2)(N-2)}}\biggr]; ...
\nonumber
\ea
Analogously, for the second line $i=1$ it follows from (\ref{diag}) that $b_{1N}=0,$ the element
$b_{1(N-1)}$ plays the role of arbitrary normalization factor, and further
\ba
&b_{1(N-2)}=b_{1(N-1)}\frac{c_{(N-1)(N-2)}}{c_{(N-1)(N-1)}-c_{(N-2)(N-2)}};\nonumber\\
&b_{1(N-3)}= .................. .
\nonumber
\ea
This procedure can be continued for $l=(N-3), (N-4), ..., l=0,$ and after that, for next lines $i=2, 3, ..., N.$
Finally, the obtained matrix $\widehat B$ is also triagonal, but with vanishing matrix elements lying under the cross diagonal.
Elements on the cross diagonal are arbitrary normalizing factors ($b_{0N},\, b_{1(N-1)}$ above). All other elements can be
compactly written
as follows\footnote{Let us mention the misprint in the upper limit of summation in analogous formula (46) in \cite{new}.}:
\be
b_{m,p}=b_{m,N-m}\cdot\biggl[\sum_{l=1}^{N-p-m}\biggl(\tau^{(m)}\biggr)^l\biggr]_{N-m,p},
\label{final}
\ee
where the $(N+1)$ triangular matrices labelled $\tau^{(m)},\,\, m=0,1,...,N$
are defined via the matrix elements of $\hat C:$
$$
\tau^{(m)}_{n,k}\equiv \frac{c_{n,k}}{c_{N-m,N-m}-c_{k,k}}.
$$
We stress that in (\ref{final}) the expression $\biggl(\tau^{(m)}\biggr)^l$
means the $l$-th power of the matrix $\tau^{(m)}.$ The repeated
index $(N-m)$ is frozen in (\ref{final})
and not summed over. This expression
allows to write all elements of the $m$-th line $b_{m,p}$ in terms
of the matrix $\tau^{(m)}$ and the arbitrary  value of the
element $b_{m,N-m}$
on the {\it crossed} diagonal. These arbitrary values can be fixed by the
normalization condition for the wave functions
$\Psi_{N-n}(\vec x)$ in (\ref{psiomega}).

\section{Shape invariance of the model}

The idea of shape invariance is one of the most essential new contributions
to the modern Quantum Mechanics made by SUSY Quantum Mechanics approach\footnote{To be honest, it should be noted that this property
was already known in the framework of well known Factorization Method of E.Schr\"odinger \cite{schr}, \cite{infeld}, but in a slightly different (and not so transparent) form.}. It was formulated originally \cite{gendenstein} in one-dimensional context, and it provided the connection between
SUSY and exact solvability of the model. Namely, all exactly solvable models were shown to obey \cite{cooper} shape invariance, which allow to solve
the models in a pure algebraic way, without solution of any differential equations. This elegant method was generalized \cite{new}, \cite{ioffe1}, \cite{periodic} to two-dimensional case where it leads only to partial (quasi-exact) solvability due to many zero modes of second order supercharges.

\subsection{First shape invariance}

It is easy to check that the shift of parameter $a \to \tilde a = a-1$ transforms $H^{(2)}$ into $H^{(1)}:$
\be
H^{(2)}(\vec x; \tilde a) = H^{(1)}(\vec x; a),
\label{shape1}
\ee
where both Hamiltonians (\ref{H}) with potentials (\ref{V}) are intertwined according to (\ref{int}). As usual \cite{new}, \cite{ioffe1}, shape invariance property allows to build the whole tower of new wave functions starting from any known wave function $\Psi^{(2)}_{n,0}(\vec x; a)$ :
\ba
&\Psi^{(2)}_{n,m}(\vec{x}; a)=Q^{-}(a)Q^{-}(a-1)...\label{shape11}\\
&...Q^{-}(a-m+1)\Psi^{(2)}_{n,0}(\vec{x}; a-m); \,\, m=1,2,3,....\nonumber
\ea
Herewith, the energy eigenvalues of $\Psi^{(2)}_{n,m}(\vec{x}; a)$ are:
\be
E^{(2)}_{n,m}(a)=E^{(2)}_{n,0}(a-m).
\label{shape12}
\ee
As a principle wave function $\Psi^{(2)}_{n,0},$ in (\ref{shape12}), one can take an arbitrary wave function $\Psi^{(2)}_n(\vec x; a)$ obtained in the previous Section (see (\ref{psiomega}) with index $n$). Its energy $E_{n,0}^{(2)}$ is given by (\ref{energies}), and therefore, the energies of states (\ref{shape11}) are:
\be
E^{(2)}_{n,m}(a)=-(A-n)^2-(A-n-m+a)^2.
\label{Enm}
\ee
One may verify that for $a=-k$
these energy levels coincide with a part of full spectrum which was
found in \cite{IKN}.

\subsection{Second shape invariance}

It was mentioned in recent paper \cite{shape2011} that among different solutions of two-dimensional SUSY intertwining relations obeying shape invariance property, some pairs are equivalent to each other up to a linear transformation of coordinates. In particular, the model considering in the present paper (the model (A5) of \cite{shape2011}) is equivalent to the model (A8) in \cite{shape2011}.

One can check that the partner Hamiltonians $\widetilde H^{(1),(2)}$ on the plane $\vec y=(y_1, y_2)$ with potentials:
\ba
&\widetilde{V}^{(1),(2)}(\vec y)=g\biggl(\frac{1}{\sinh^{2}(\lambda y_{2})}-\frac{1}{\cosh^{2}(\lambda y_{1})}\biggr)+\nonumber\\
&+\frac{\alpha\lambda(2b\mp1)\sinh(\lambda y_{+})-2\lambda^{2}b(b\mp1)+\alpha^{2}/2}{\cosh^{2}(\lambda y_{+})}+ \nonumber\\
&+\frac{\alpha\lambda(2b\mp1)\sinh(\lambda y_{-})-2\lambda^{2}b(b\mp1)+\alpha^{2}/2}{\cosh^{2}(\lambda y_{-})};\label{2.1}\\
&y_{\pm}\equiv y_1 \pm y_2
\nonumber
\ea
are solutions of intertwining relations of the form (\ref{int}):
\be
\widetilde{H}^{(1)}(\vec y)\widetilde{Q}^{+}(\vec y)=\widetilde{Q}^{+}(\vec y)\widetilde{H}^{(2)}(\vec y) \label{22.1}
\ee
with the supercharges\footnote{We use everywhere in the present paper the same definitions of coefficient functions $C_{\pm}$ as in \cite{shape2011}, \cite{IKN}. They differ from definitions of $C_{\pm}$ in all earlier papers by the factor $1/4.$}:
\ba
&\widetilde{Q}^{\pm}(\vec{y})=4\partial_{y_{+}}\partial_{y_-}\pm4\widetilde C_{+}(y_+)\partial_{y_-}\pm4\widetilde C_{-}(y_-)\partial_{y_+}+\widetilde B(\vec{y}); \nonumber\\
&\widetilde B(\vec{y})=4\widetilde C_{+}(y_+)\widetilde C_{-}(y_-)+\widetilde f_{1}(y_1)+\widetilde f_{2}(y_2);\nonumber\\
&\widetilde C_{+}(y_+)=b\lambda \tanh(\lambda y_{+})+\frac{\alpha}{2\cosh(\lambda y_{+})};\nonumber\\
&\widetilde C_{-}(y_-)= b\lambda \tanh(\lambda y_{-})+\frac{\alpha}{2\cosh(\lambda y_{-})};\nonumber\\
&\widetilde f_{1}(y_1)=\frac{g}{\cosh^{2}(\lambda y_{1})};\quad \widetilde f_{2}(y_2)=\frac{g}{\sinh^{2}(\lambda y_{2})}. \nonumber
\ea
According to \cite{shape2011}, this model is shape-invariant:
$\widetilde{H}^{(2)}(\vec{y},b-1)=\widetilde{H}^{(1)}(\vec{y},b),$
and therefore, if one knows any (principal) wave function $\widetilde\Psi^{(2)}_{n,0}(\vec y)$ of $\widetilde H^{(2)}(\vec y),$
then the whole tower of wave functions can be built due to shape invariance:
\ba
&\widetilde{\Psi}^{(2)}_{n,m}(\vec{y}; b)=\widetilde Q^{-}(\vec y; b)...\widetilde Q^{-}(\vec y; b-m+1)
\widetilde{\Psi}^{(2)}_{n,0}(\vec{y}; b-m);\nonumber\\
&m=1,2...;\,\, \widetilde{E}_{n,m}(b)=\widetilde{E}_{n,0}(b-m).\label{2222.2}
\ea

At the first sight, the model (\ref{2.1}) has no relation to the Scarf II model (\ref{V}). But as it was noticed in \cite{shape2011},
they are related by substitution of coordinates: $\, x_{+} \equiv y_{1};\, x_{-} \equiv y_{2},$
since (\ref{2.1}) can be rewritten as:
\ba
&\widetilde{V}^{(1),(2)}(\vec x)=g(\frac{1}{\sinh^{2}(\lambda x_{-})}-\frac{1}{\cosh^{2}(\lambda x_{+})})+
\label{3.2}\\
&+\frac{\alpha\lambda(2b\mp1)\sinh(2\lambda x_{1})-2\lambda^{2}b(b\mp1)+\alpha^{2}/2}{\cosh^{2}(2\lambda x_{1})}+\nonumber\\
&+\frac{\alpha\lambda(2b\mp1)\sinh(2\lambda x_{2})-2\lambda^{2}b(b\mp1)+\alpha^{2}/2}{\cosh^{2}(2\lambda x_{2})}.\nonumber
\ea
If one identify the coupling constants in (\ref{3.2}) and (\ref{V}) as follows:
\ba
&g=\lambda^{2}a(a+1);\quad \alpha\lambda (2b+1)=B(A+\lambda );\label{2.2}\\
&\alpha^{2}-4\lambda^{2}b(b+1)= (B^2-A^2-2A\lambda ),\nonumber
\ea
it becomes evident, that the second superpartners are proportional: $H^{(2)}(\vec x)=2 \widetilde H^{(2)}(\vec x),$
and thus, the Hamiltonian $H^{(2)}(\vec x)$
satisfies simultaneously two intertwining relations - with $H^{(1)}$ and $\widetilde H^{(1)} :$
\be
H^{(1)}(\vec x; a, b) \div H^{(2)}(\vec x; a, b) \equiv
2 \widetilde H^{(2)}(\vec x; a, b) \div 2 \widetilde H^{(1)}(\vec x; a, b),
\label{3.8}
\ee
where the sign $\div $ means intertwining, and both shape invariance parameters $a, b$ of potentials are written explicitly.

For the particular case of $2\lambda =1,$ the solution of system (\ref{2.2}) is $b=A;\, \alpha =B,$ and therefore, the energies
in (\ref{2222.2}) coincide with levels (\ref{Enm}).
By means of laborious but elementary calculations, one can check straightforwardly, that
\ba
\widetilde Q^{-}(\vec x; a, b)Q^{-}(\vec x; a, b-1)=Q^{-}(\vec x; a, b)\widetilde Q^{-}(\vec x; a-1, b). \label{3.9}
\ea
Similarly to the situation \cite{double} for the Morse model, this identity means that starting from the arbitrary
principle state $\Psi^{(2)}_{0,0}(\vec x; a, b),$ one may combine both shape invariances (in parameter $a,$ and parameter $b$)
to obtain wave functions  $\Psi^{(2)}_{n,m}(\vec x; a, b).$ In other words, one can move by different paths in the
plane $(a, b)$ of parameters.

\acknowledgements

E.V.K. is indebted to the non-profit foundation "Dynasty" for financial support.

\end{document}